\documentclass[10pt,conference]{IEEEtran}
\IEEEoverridecommandlockouts
\usepackage{url}
\usepackage{cite}
\usepackage{amsmath,amssymb,amsfonts}
\usepackage{algorithmic}
\usepackage{graphicx}
\usepackage{textcomp}
\usepackage{xcolor}
\usepackage{subfigure}

\begin{document}

\title{Experiences and insights from using Github Classroom to support Project-Based Courses}

\author{\IEEEauthorblockN{Maria Augusta Nelson}
\IEEEauthorblockA{Dept. of Software Engineering and Information Systems \\
Pontifical Catholic University of Minas Gerais\\
Belo Horizonte, Brazil \\
guta@pucminas.br} 
\and
\IEEEauthorblockN{Lesandro Ponciano}
\IEEEauthorblockA{Dept. of Software Engineering and Information Systems\\
Pontifical Catholic University of Minas Gerais\\
Belo Horizonte, Brazil\\
lesandrop@pucminas.br}
}

\maketitle

\begin{abstract}
This work presents an approach for using GitHub classroom as a shared, structured, and persistent repository to support project-based courses at the Software Engineering Undergraduate program at PUC Minas, in Brazil. We discuss the needs of the different stakeholders that guided the development of the approach. Results on the perceptions of professors and students show that the approach brings benefits. Besides the lessons learned, we present insights on improving the education of the next generation of software engineers by employing metrics to monitor skill development, verifying student work portfolios, and employing tooling support in project-based courses.
\end{abstract}
 

\section{Introduction}

Project-based learning~\cite{doi:10.1177/1365480216659733} has been used worldwide as an active methodology~\cite{Freeman8410} to provide the students with a more realistic and professional experience. In the software engineering undergraduate program at PUC Minas in Brazil, we make extensive use of project-based courses. During each semester of the program, students have to develop a software application \cite{Nelson:2017}. In general the projects are a way to motivate the students to take responsibility for their learning process. They challenge themselves through the process and go deeper and further than in traditional courses. Furthermore, since this generation is more socially aware they are keen to solve problems brought by the surrounding community. 

In each semester, the projects have different goals and are always related to the other courses that students are taking in the same semester. These other courses are the anchors for the projects. Students work in groups that vary in size from 3 to 6 depending on the project and on the specific course. Releases occur every week or every other week depending on the course. Usually they like to collaborate and improve these skills. Because of this team work, we have always motivated the students to find repository tools where they could share their work, collaborate and keep track of the versions of the system under development. During the project, students would share their work with their supervising professor so that they can provide feedback and directions on how to continue, as well as suggest improvements. 

From the point of view of the professors, the shared repositories were only a way to visualize student work. Each group used a different repository according to the familiarity that they had with such tools. Sometimes, even inside a single team it was hard to achieve consensus about which tool to use, and we have awkward decisions such as two repositories in different tools, one for the backend and another for the frontend. By observing these not-so-good practices, the supervising professors started to discuss what would be a better way to improve this situation.

Some of these projects have real clients. Two project courses are officially considered extension courses, meaning the students work directly with the community surrounding the university to provide this community with a concrete return of the application of the students' knowledge. At the end of the semester, some of these clients and stakeholders want to further develop and elaborate these projects. They also need to have access to these repositories. Furthermore, some of these projects span more than one semester, done by different teams. The need to have more control over these repositories to guarantee their longevity became apparent. 

Considering the needs of the program and the five professors who coordinate it, it is necessary to keep track of the projects to register what students are doing, to evaluate how the project-based courses are evolving, and to showcase the best projects. Every semester the best projects are published in the program's website and in a magazine describing the clients' problems and software solutions the teams came up with.  

All those different needs and perspectives result in a challenge to manage projects every semester and their evolution overtime. Currently, we have approximately 60 projects per semester, with 287 enrolled students, and 13 supervising professors. Seeking a software approach that allows us to manage learning, participation and meet stakeholders needs, we developed an approach based on the GitHub Classroom tool. In the next sections, we describe the proposed approach (Section~\ref{sec:approach}), results of experience in its use (Section~\ref{sec:experience}), and the road ahead, analyzing the new insights on metrics for monitoring their learning, on student work portfolios, and on tooling support for project-based courses (in Section~\ref{sec:ahead}).

\section{A GitHub Classroom Approach}
\label{sec:approach}

GitHub Classroom\footnote{Available at: https://classroom.github.com/ (Accessed: 01 January 2021)} is a tool that provides a dashboard for standardizing, organizing, and managing student' code repositories and grading work automatically. The GitHub Classroom is integrated with the GitHub platform\footnote{Available at: http://github.com/ (Accessed: 01 January 2021)}. Students and professors must have an account on this platform. In the GitHub Classroom terminology, a {\it classroom} is composed of a set of students, a set of professors and teaching assistants. When created, a classroom is linked to an {\it organization} on the GitHub platform. Classrooms of all courses in the undergraduate program are associated with the same organization. The inclusion of students in the classroom can be done automatically through an integration with the Learning Management Systems. 

Professors and students act on GitHub Classroom as follows. Professors can create classrooms and, within those classrooms, create {\it assignments} for students. An assignment is usually a coding task, which can be performed by a student individually or by a group of students. For each student/group who accepts an assignment, GitHub Classroom creates a {\it code repository} automatically.  Depending on the configuration made by the professor, the repository can be public or private, and can also have a standard structure of folders and files based in a predefined template. The student receives the link to the repository created by GitHub Classroom. All student's repositories associated with the classroom are maintained in the organization. Students can act in their repositories the same way they do in any other repository on the GitHub platform, by performing typical operations such as: clone, commit, pull, and push. They can also document and manage their projects by creating {\it wiki pages} and {\it project boards} made up of issues, pull requests, and notes.

The novelty of this approach is using GitHub classroom not just on a course, but in all project-based courses within the scope of the undergraduate program in a shared,  structured, and institutionalized way. The GitHub organization belonging to the  program is central. An organization is a namespace where all student's repositories exist. It provides tools to manage subgroups of people with shared ownership of repositories. While the classroom is specific to a course and is archived at the end of the semester, the repositories stored in the organization persist active overtime. They can be maintained by other students, teachers and provided to clients or other stakeholders. Besides each professor having the possibility to analyze the learning outcomes of his/her students over the semester, an analysis of all students enrolled in the undergraduate program is also possible. From the organization, coordinators of the program can take a snapshot of classwork produced by students in all courses in each semester.

\section{Report of the first experience}
\label{sec:experience}

On the first semester of 2020 we decided to run a pilot of our approach based on GitHub Classroom in one of the eight project-based courses. We chose the course on the fourth semester of the curriculum because of a few reasons: 1) students are mature enough and really understand the good practice of sharing a common repository among the team members; 2) in this course there are real clients for the projects and at the end students have to hand in the complete project to the client; and, 3) in this course a project in one semester is usually continued by another team in the following semester. 

Considering all these reasons, starting with the fourth semester course was a good choice. The classrooms were created by the owner of our own organization in GitHub, one of the professors who also coordinates the program. Besides creating the classrooms, he also gave owner permissions to the supervisors. Following the approach discussed in Section~\ref{sec:approach}, students accepted the assignment and received a link to enter the repository automatically generated by GitHub classroom. Each team used its repository throughout the semester. This pilot experience was perceived as being successful by the supervising professors.

In the second semester of 2020 we decided to adopt our GitHub Classroom approach for all eight project-based courses. We had a meeting in the beginning of the semester where we explained to all supervising professors the dynamics of the tool. We also created a repository template to be used for all courses. Supervising professors could add to the template structure for each of their courses but they had to follow a primitive structure. In this structure we guarantee folders and files to be organized in such a way as to ensure that at the end of the semester we would find every information we may need on these projects, and that students understand the organization of the different types of artifacts produced in the course.

In this section, we detail the results of the experience by describing professor' perceptions, student' perceptions and lessons learned.

\subsection{Professor Perceptions}

We had a meeting at the end of the second semester of 2020 to hear how the faculty felt about using GitHub classroom. All 13 supervising professors who participated in the experience and used the proposed GitHub classroom approach attended the meeting. In general they reported that they want to learn more about the tool to be able to use more features in the projects, and to help them with marking, and providing feedback to the students. A recurring demand in the faculty is the definition of a set of metrics measured directly from the student's activity log in the repositories and that supports the assessment process.

\subsection{Student Perceptions}

We also conducted a survey with students to see what was their impression of using the repository environment set up in the course. We had 287 students enrolled in project-based courses. They all received the survey to answer and we received 58 answers back (approximately $20\%$).

From the 58 respondents, 44 ($75.86\%$) thought the experience was positive, 3 ($5.17\%$) thought the experience was negative, and 11 (18.97\%) had a neutral opinion. About the work in teams using GitHub as a repository, 42 ($72.41\%$) students said that was satisfactory, 12 ($20.69\%$) said it brought more complexity to the team and project dynamics, and 4 ($6.90\%$) said it was indifferent.

\begin{figure*}[bht]
    \centering
    \includegraphics[width=11.1cm]{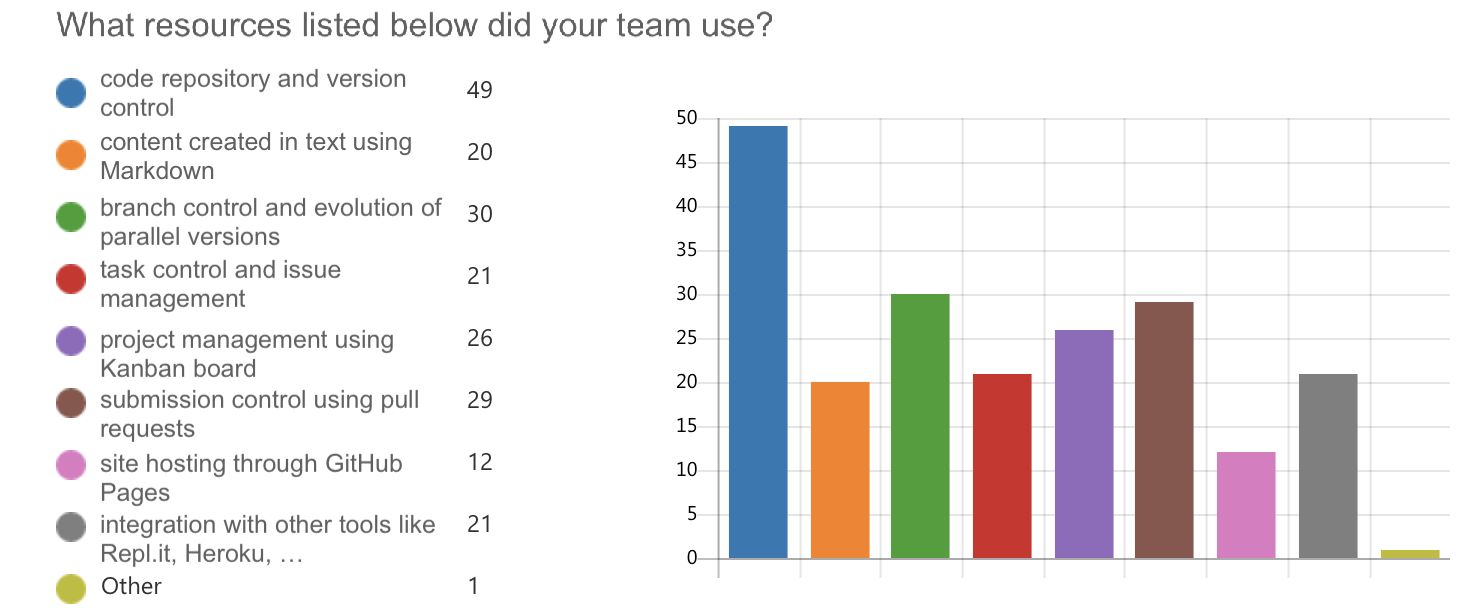}
    \caption{Resources used by the teams. Total of 58 respondents, each being able to select more than one option.}
    \label{fig:resources}
\end{figure*}

\begin{figure*}[htb]
    \centering
    \subfigure[Level of Commitment ]{\includegraphics[width=8.3cm]{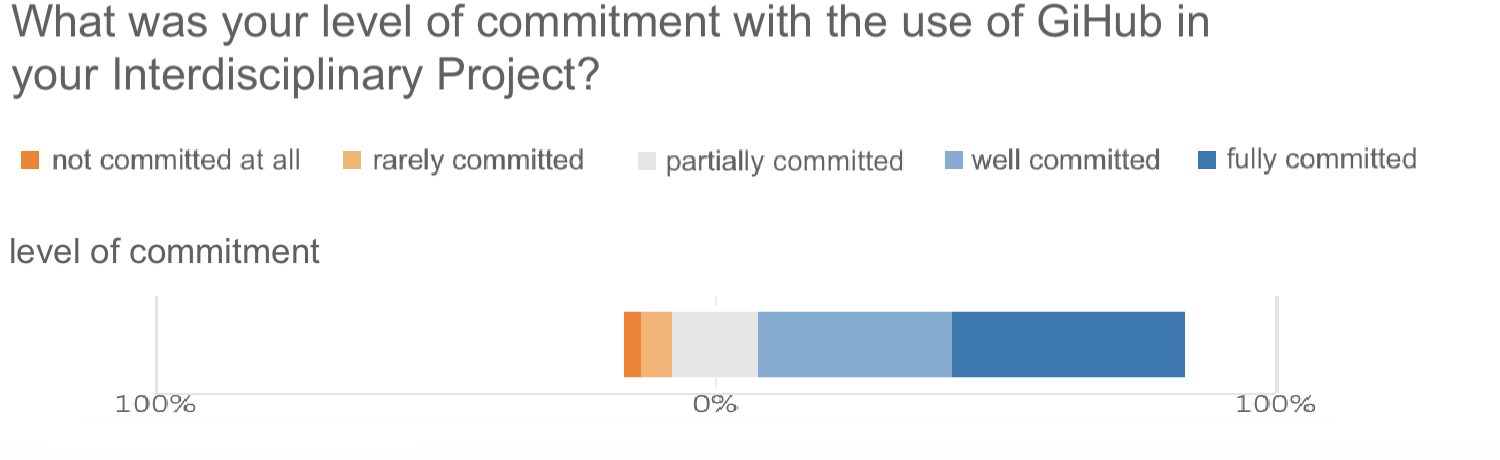}}
    \subfigure[Interest in Learning about GitHub]{ \includegraphics[width=8.3cm]{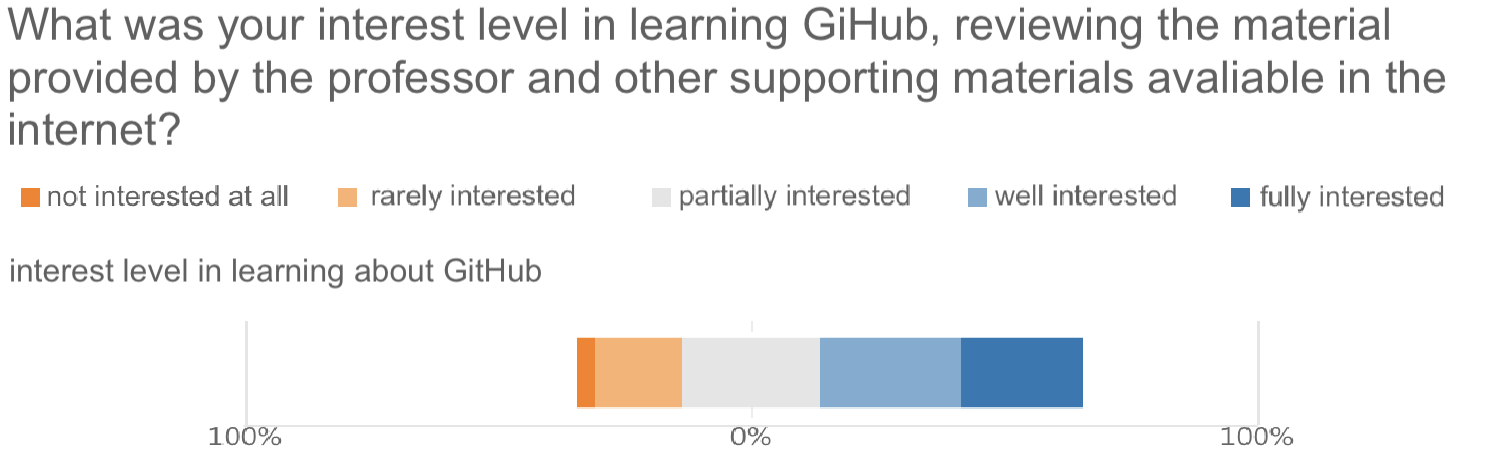}}
    \caption{Students' self-reported commitment and interest. Total of 58 respondents, each being able to select only one option.}
    \label{fig:commitmentinterest}
\end{figure*}

\begin{figure*}[hbt]
    \centering
    \subfigure[Difficulties, being each student able to select more than one option.]{\includegraphics[width=8.3cm]{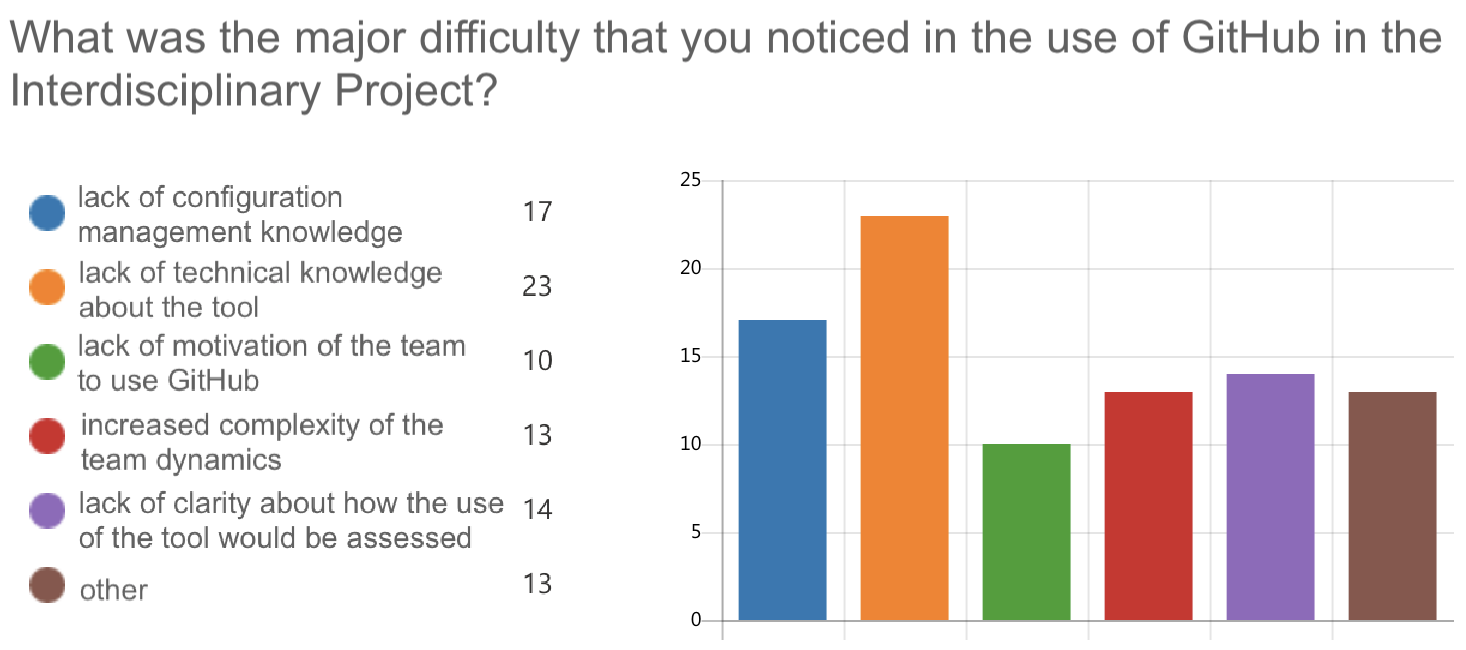}}
    \subfigure[Benefits, being each student able to select more than one option.]{\includegraphics[width=8.3cm]{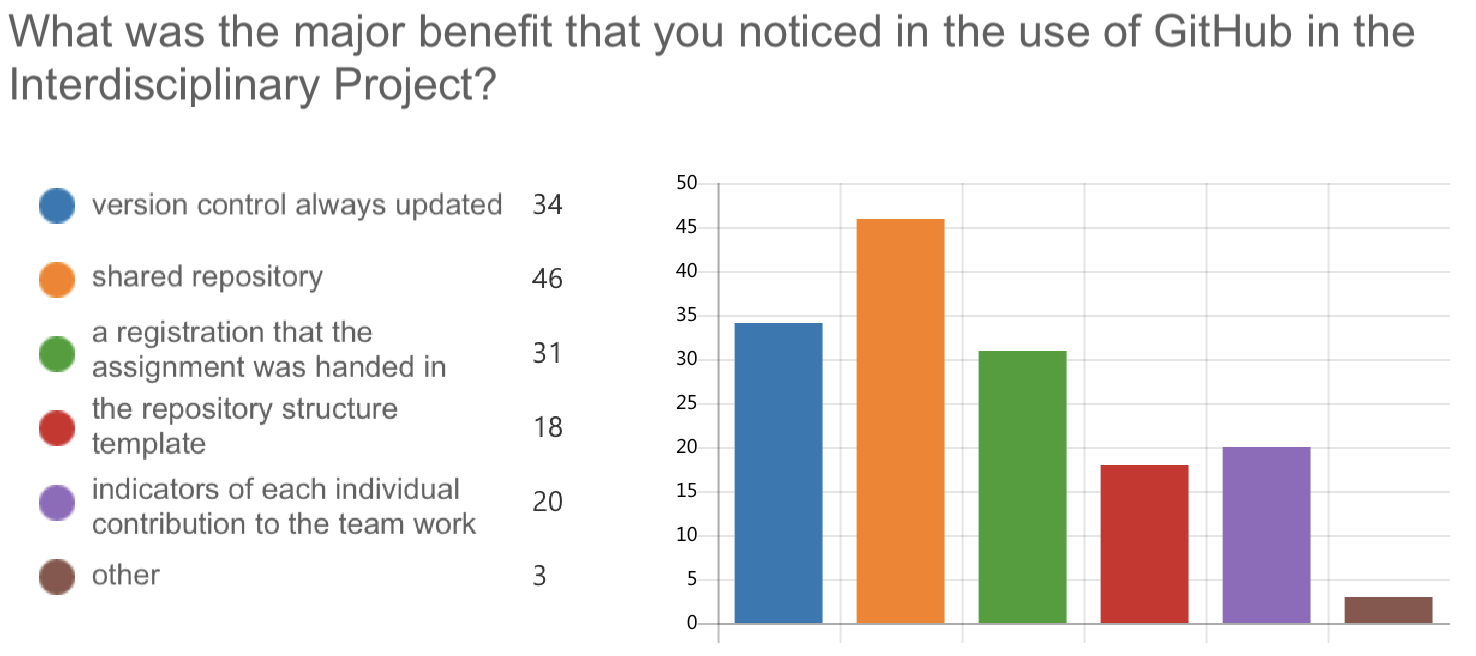}}
    \subfigure[Recommendation. Each student is able to select only one option.]{\includegraphics[width=8.2cm]{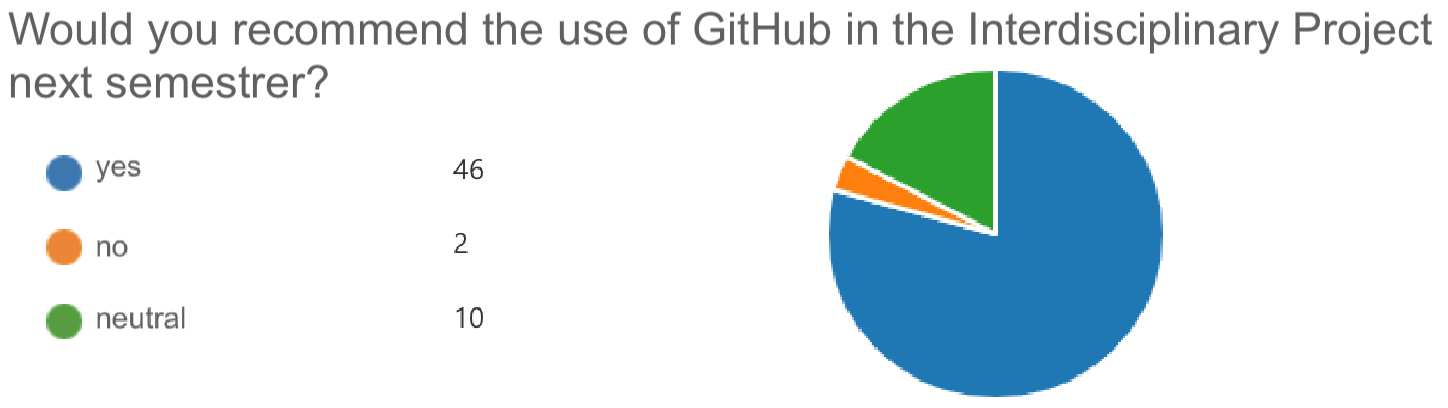}}
    \caption{Students' perceptions and recommendations. Total of 58 respondents.}
    \label{fig:perceptions}
\end{figure*}

Asked to provide a general score to the experience, the averaged score was \textit{4.3 out of 5}, which we think is very good for a first experience.  We also evaluate more deeply the resources used by the teams (Fig.~\ref{fig:resources}), students commitment and interest in using our GitHub-based approach (Fig.~\ref{fig:commitmentinterest}), students' perceived benefits, difficulties, and recommendations (Fig.~\ref{fig:perceptions}). 

Figure~\ref{fig:resources} shows a histogram of the resources used by the teams during the project-based course with our GitHub Classroom approach. Most used resources were code repository and version control, branch control, and evolution of parallel versions. This is expected because our infrastructure has just been implemented and there is still a lot to be explored, in terms of the resources provided by GitHub classroom, GitHub organization and repositories, as we detail in Section~\ref{sec:ahead}.

The students assessed their level of interest and commitment on a 5-points semantic differential scale ranging from not {\it committed/interested at all} to {\it fully committed/interested} (Figure~\ref{fig:commitmentinterest}). Overall, the results show that students are more likely to be very committed than  not committed, with 76.9\% indicating well or fully committed. The equivalent tendency was observed for the level of interest, with $54,2\%$ indicating well or fully interested.

Finally, Figure~\ref{fig:perceptions} shows students' perceived benefits, difficulties, and overall recommendations. The major difficulty was lack of technical knowledge about the tool ($39.65\%$ of students). The major benefit was the shared repository used by the team in the organization ($79\%$ of the students). More than three quarters of students ($79\%$ of the students) recommended us to stick to the approach in the following semesters.

In addition to these questions, students also had a space where they could write about the experience in general and provide suggestions. Some suggestions include: ``we would like to use Markdown to prepare the project documentation instead of Word''; ``I would rather use the repository tool that my team finds most appropriate''; ``My team was not able to integrate our repository directly with Heroku and that made it much harder to deploy in Heroku.''

\subsection{Lessons Learned}

From professors, to students, to coordinators and clients in the community, everyone saw an improvement from previous experiences. In general the majority found that the experience was good. From theses results we could observe the following factors to improve: 1) students felt the need for more training with the tool; 2) students need to understand more about configuration management to take full advantage of this proposal; and 3) students need to understand the benefits in using the tool as soon as they start their projects. 

From these students answers we have decided to: 1) allow and motivate students to use Markdown for the project documentation which will allow professors to understand the contribution of the team to the project documentation in the same manner as we can do with code; 2) motivate the students to ensure they understand the benefits of using the tool, and to prepare more training; and 3) to make the repositories public to allow for the integration with other tools and environments.

\section{The Road Ahead: New Requirements and Improvements}
\label{sec:ahead}

Using code versioning platforms and repository management in the teaching-learning context has become increasingly popular in teaching software engineering~\cite{Feliciano:2016,Parizi2018,Angulo2019,Hsing:2019}, but also in teaching other skills~\cite{Fiksel2019}. In this work, we {\it use GitHub classroom not just on a course, but in all project-based courses in the undergraduate program in a shared, structured, and  institutionalized way.} This approach allows us to provide students with project-based learning closer to the software development approach practiced in the industry~\cite{garousi:2020}. 
The first experiences provide us with insights on how this can evolve to enable new innovations in the teaching-learning process of software engineering. In this section, we focus on discussing these insights.

When applying a project-based learning methodology to courses with many students, monitoring the gain in knowledge and the development of skills and attitudes by each student is challenging. For example, a student may hide behind his team and it is hard to monitor the contributions of each student. This monitoring is carried out by professors to conduct formative assessment, provide individual feedback to each student on each team, and foster students' self-reflection~\cite{Baltes2018}. A standardization and centralization based on repositories, following project management practices, may {\it enable the automatic extraction of learning metrics and indicators to support analysis of knowledge, skills, and attitudes}. There are some efforts in this direction, focusing on measuring engagement, collaboration and quality of contribution to the team~\cite{Makiaho2015,Parizi2018}, mitigating possible inaccuracies and attempts at cheating~\cite{Tushev2020,Mohammad2020,Kun:2020}. However, there is a lack of convergence and standardization for metrics and learning indicators suitable for each stage of student training. On an undergraduate program in which each semester the student works on a project, changing the scope and increasing the complexity of the project, the definition of metrics and learning indicators suitable for each stage is relevant. To this end, it is necessary to {\it establish a theoretical framework covering knowledge, skills and attitudes associated with the students' maturity in each stage of a sequence of project-based courses}. 

In addition to the other skills that are developed in the projects, but that are not directly associated with coding tasks, our approach of integrating GitHub Classroom, GitHub organizations and the students' GitHub  in the different projects-based courses allows the student to obtain, in an iterative and incremental way, experiences with software licensing, collaboration in software development teams, configuration management, and continuous integration. {\it An educational tooling support yet to be fully explored and exploited is the integration with other apps/bots/extensions that can be used with GitHub repositories}, such as automating quality assurance procedures and software deployment activities. The use of this sort of tool is increasingly common in the industry ~\cite{lebeuf2017software,Lebeuf:2019} and is a demand we have received from students. They discover apps/bots developed through frameworks like ProBot\footnote{Available at: https://probot.github.io/apps/ (Accessed: 07 January 2021)} that they would like to use in their projects and request our authorization to use them in the organization's repositories. We believe there is room for using, testing, and developing tools to support software engineering activities within this context, directly as a tool for the GitHub platform or as an external tool integrated via the application programming interface (API) provided by that platform. 

Our approach for managing student's repositories in project-based courses allows the {\it creation of the students' portfolio verified by professors during the undergraduate program}. 
Such a portfolio is of academic and professional importance to students~\cite{cheng2019innovative,olaniyi2020survey}. Nowadays, GitHub profiles act as a portfolio signaling technical expertise and experience of developers~\cite{Feliciano:2016,Hsing:2019,Wang:2020}. It is a relevant factor in the hiring process at many companies. In this way, the student's skills are demonstrated in the projects in which she/he worked. Of course, as the learning process is continuous and incremental, a portfolio that describes the skills already acquired can also point out which new skills and experiences are yet to be developed. This can become a new tool and parameter of analysis to be considered by both students and professors when defining the next steps on the learning path.

To summarize, we advocate that: (1) using tooling support is essential in the teaching-learning process in Software Engineering when students, professors, program coordinators, and real clients are involved; and (2) building a shared, structured, and persistent repository in project-based courses makes it possible to integrate these stakeholders and support education by employing metrics to monitor skill development, conducting formative assessment, providing individual feedback in teamwork, and verifying student work portfolios.

\section{Author Profiles}

{\bf Maria Augusta Nelson} received her Ph.D. degree in Computer Science from the University of Waterloo, Canada, in 2003. She is an Associate Professor at the Pontifical Catholic University of Minas Gerais (PUC Minas), in Brazil, where she has been teaching Software Engineering, Software Testing, and Software Design courses since 2004. She was part of the team that conceived the Software Engineering Undergraduate Program and is now in charge of the program.  She leads the Structuring Teaching Core, which is a group of professors  responsible for conceiving, checking, questioning, and improving the pedagogical structure of the program. Her research interests are in computer education, focusing in software engineering education, requirements engineering and curriculum design involving also distance and continuing education. She is a member of the Brazilian Computer Society, participating in the Computer Education Interest Group (GIEC). She has served in the program committee of the IEEE/ACM 1ST International Workshop On Software Engineering Curricula for Millennials (SECM), and the second International Workshop On Software Engineering Education for Millennials (SEEM). Further information is available on her ORCID at https://orcid.org/0000-0002-1151-1362.

{\bf Lesandro Ponciano} received a Ph.D. degree in Computer Science from the Federal University of Campina Grande, Brazil, in 2015. He is an Associate Professor at the Pontifical Catholic University of Minas Gerais (PUC Minas), in Brazil, where he has been teaching Requirements Engineering, Software Testing, and Human-Computer Interaction courses since 2016. In the Software Engineering program of PUC Minas, he is also a member of the Structuring Teaching Core, which is a group of professors who are responsible for conceiving, checking, questioning, and improving the pedagogical structure of the program. His research interests are in computer education, human-computer interaction, and software engineering, focusing on engagement, credibility, coordination, and learning analysis. He is a member of the Brazilian Computer Society, participating in the Computer Education Interest Group (GIEC). He is certified as a GitHub Campus Advisor\footnote{Available at: https://education.github.com/teachers/advisors Accessed: 07 January 2021)}, which is a program part of the GitHub Education. Further information is available on his ORCID at https://orcid.org/0000-0002-5724-0094.

\bibliographystyle{IEEEtran}
\bibliography{refs}

\end{document}